\documentclass[lettersize,journal]{IEEEtran}
\usepackage{amsmath,amsfonts}
\usepackage{algorithmic}
\usepackage{algorithm}
\usepackage{array}
\usepackage[caption=false,font=normalsize,labelfont=sf,textfont=sf]{subfig}
\usepackage{textcomp}
\usepackage{stfloats}
\usepackage{url}
\usepackage{verbatim}
\usepackage{graphicx}
\usepackage{cite}
\usepackage{siunitx}
\usepackage{lineno}
\begin{document}

\title{Thin single-crystal perovskite detector for high-energy  charged particles}
\author{
Z. Chubinidze,
M. Auf der Maur,~\IEEEmembership{Member,~IEEE},
F. Matteocci,
I. Viola,
J. Endrizzi,
L. Pancheri,~\IEEEmembership{Senior Member,~IEEE},
A. Khan,
G. Papalino,
G. Felici,
A. De Santis,
G. Tinti,
M. Testa\thanks{M. Testa is the corresponding author (e-mail: marianna.testa@lnf.infn.it).}
\thanks{Zaza Chubinidze and Amir Khan are with INFN-LNF, 00044 Frascati (RM), Italy, and with CHOSE (Centre for Hybrid and Organic Solar Energy), Department of Electronic Engineering, University of Rome “Tor Vergata”, via del Politecnico 1, Rome, 00133 Italy, e-mail: zaza.chubinidze@lnf.infn.it.}
\thanks{Marianna Testa, Giuseppe Papalino, Giulietto Felici, Antonio De Santis and Gemma Tinti are with INFN-LNF, 00044 Frascati (RM), Italy}
\thanks{Matthias Auf der Maur and Fabio Matteocci are with  CHOSE (Centre for Hybrid and Organic Solar Energy), Department of Electronic Engineering, University of Rome “Tor Vergata”, Via del Politecnico 1, Rome, 00133 Italy, and with INFN-LNF, 00044 Frascati (RM), Italy,}% <-this % stops a space
\thanks{Ilenia Viola is with CNR-NANOTEC, Piazzale Aldo Moro, 5 00185 Roma (RM) Italy and with INFN-LNF, 00044 Frascati (RM), Italy}% <-this % stops a space
\thanks{Jacopo Endrizzi and Lucio Pancheri are with the Department of Industrial Engineering, University of Trento, 38123 Trento, Italy and INFN-TIFPA, 38123 Trento, Italy}% <-this % stops a space
\thanks{
Funded by the European Union under NextGenerationEU through the PNRR program of the Italian Ministry of University and Research (MUR), PRIN 2022, project 'HyPoSICX', Prot. No. 2022LWHCWY.}% <-this % stops a space
}

\markboth{Journal of \LaTeX\ Class Files,~Vol.~14, No.~8, August~2021}%
{Shell \MakeLowercase{\textit{et al.}}: A Sample Article Using IEEEtran.cls for IEEE Journals}

\IEEEpubid{0000--0000/00\$00.00~\copyright~2021 IEEE}

\maketitle
%\linenumbers 
\begin{abstract}

The organometal halide perovskites (OMHP) semiconductors have shown recently a strong potential as radiation detectors, beside the well-known success in photovoltaics and as photo-detectors. Many studies have been published on X-rays detection, and a few studies about detection of $\alpha$-, $\beta$-particles and protons. Less literature is present for high energy charged particles.
OMHP-based devices for  tracking and real-time monitoring for high energy particles may offer many advantages. OMHPs can be  directly grown on pixelated electronics, even on curved substrates,  without the need of complex and expensive bump-bonding procedures. Moreover, OMHPs have shown self-healing features after radiation exposure, which makes them attractive for high-flux applications.

In this paper we report
a device based  on a thin single OMHP crystal, $\sim$150 $\mu$m thick, directly grown on a patterned substrate through dewetting technique, 
able to detect  high-energy charged particles in a high dynamic range of incident fluxes. A dedicated electronics circuit has been developed to match the expected time characteristics of the OMHP crystals.
This is the first demonstration of thin OMHP single crystals being able to detect  high energy charged particles of  hundreds of MeV.

\end{abstract}


\section{Introduction}
There is a steadily growing interest in  novel thin, high-sensitivity and flexible solid-state  devices  for radiation detection characterized by low manufacturing costs and easy-processing. 
The organometal halide perovskites (OMHP) semiconductors are attractive as sensor for those devices since  they have a strong stopping power, defect-tolerance, large mobility lifetime ($\mu \tau$) product, tunable bandgap, and they can be grown through low-cost solution processes~\cite{Sakhatskyi2025,Liu2024,Yihui2022}. Particularly attractive is the possibility to directly grow the OMHP on substrates~\cite{Viola2023}, even curved~\cite{Li2022,GU20232666}, that integrate the electronics. This feature  allows to avoid the expensive and complicated flip-chip process needed for hybrid pixel detectors, like silicon sensors bump-bonded on ASIC chips.

The OMHPs  also present intriguing self-healing properties after radiation exposure~\cite{Huang2023,Afshari2023,Kirmani2024,Yin2026}. 
They could therefore have potential as devices for monitoring charged particles in harsh environments, for example in medical applications such as FLASH therapy. In particular for this therapy, there is a strong interest for very-high-energy electron  beams, ranging from 50 to  400 MeV~\cite{qubs9040029}.  

In~\cite{Testa2024} the authors reported a device based on a 1 mm 
thick  methylammonium lead bromide (MAPbBr3) bulk-crystal that is sensitive to single minimum ionizing particles (MIPs) and linear up to 10$^4$ particles per 15 ns bunch.
However, bulk crystals cannot be grown directly on the surface of readout electronic platforms, such as CMOS ASICs or TFT backplanes. Moreover, given their thickness they require high voltage bias. Also, time performance and radiation hardness may be limited by the large distance between electrodes.

Some studies have been published on thin perovskite-based devices for charged particle beam monitoring.
In~\cite{Bruzzi2023}, a 0.4 - 1 $\mu$m-thick film of  CsPbCl$_3$ has been deposited on plastic flexible substrates equipped with interdigitated electrodes. The device is sensitive to protons in the range 100–228 MeV. In~\cite{Fratelli2024}  a flexible and large area device based on 2D hybrid perovskite thin films has been realized to detect a 5 MeV proton beam with fluxes of 10$^8$ - 10$^{11}$ s$^{-1}$cm$^{-2}$.
In both cases, the devices are based on a continuous read-out of the current generated inside the perovskite rather than the  recoding of individual pulses. Generally, the principle of the current readout is related to the time characteristics of proton beams that may be considered almost continuous.
Thin perovskite-based devices for real-time monitoring of high energy electrons in a beam with a pulsed time structure have not been reported yet.
In this article, we present a perovskite detector based on a thin OMHP crystal suitable for monitoring high-energy electrons. Section II describes the fabrication procedure and structure of the devices. Section III discusses their preliminary electrical and optical characterization. Sections IV and V present the design and calibration of a readout electronic circuit suitable for amplifying the signal from single events. Finally, section VI discusses the results obtained with a detector characterization at a test beam with high-energy electrons.
\section{Crystal Growth and Device Fabrication}
\label{sec:device}
\begin{figure*}[!t]
\centering
\includegraphics[width=1.5\columnwidth]{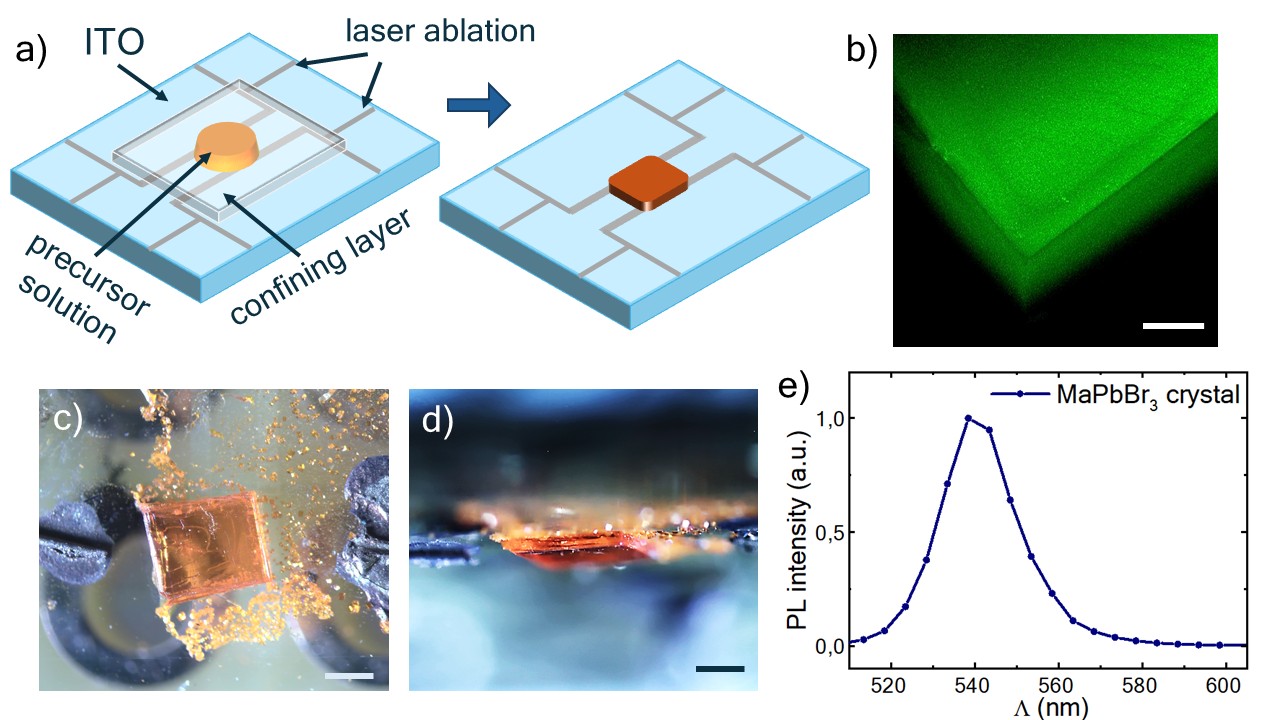}
\caption{(a) Schematic illustration of the growth process of a MaPbBr$_3$ crystal via solution dewetting directly on a patterned ITO conductive substrate. The different device components and the main stages of the dewetting-driven crystallization process are highlighted.
(b) Confocal microscopy image of a representative region of the MaPbBr$_3$ crystal obtained using the process shown in (a).
(c) Optical microscope image of the final photodetector, featuring a square-shaped MaPbBr$_3$ crystal with lateral dimensions larger than 1 mm and top contacts deposited at the crystal interface.
(d) Side-view optical image of the same photodetector, allowing evaluation of the crystal thickness, measured to be approximately 150 $\mu$m.
(e) Spatially resolved photoluminescence spectrum acquired by confocal measurements directly on the MaPbBr$_3$ crystal. Scale bars: (b) 50 $\mu$m; (c,d) 500 $\mu$m.}
\label{fig:FigDewetting}
\end{figure*}

Macroscopic MAPbBr$_3$ crystals were synthesized directly on a patterned indium tin oxide (ITO) substrate by exploiting a solution-based dewetting-assisted growth approach, using an elastomeric confining layer (Fig.\ref{fig:FigDewetting} a).
The dewetting approach provides precise control over the supramolecular organization of molecular precursors from solution, enabling the formation of high-quality microcrystals. By modulating spatially confined, super-saturated regions, dewetting enhances local ordering, amplifies interfacial instabilities, and directs the anisotropic assembly of building blocks~\cite{Viola2023,Deng2015,Mao2017,Yang2018}.
Here, controlled dewetting of the precursor solution on patterned ITO surface induces spatial confinement of the liquid phase, promoting localized nucleation and guided crystal growth. The perovskite solution has been prepared by mixing PbBr$_2$ and MABr perovskite precursors at 2M concentration in DMSO solvent.
In our experimental configuration, dewetting is driven by the progressive evaporation of a small volume of precursor solution deposited on a non-wetting substrate. Evaporation initiates at the liquid–air interface and induces instabilities within the thinning liquid film, which trigger nucleation events and promote precursor redistribution and constructive self-assembly. The onset of perovskite nucleation is preferentially observed at boundaries between regions of different wettability, which act as energetically favorable sites where the total free energy of the system is minimized. Dewetting promotes localized ordering and anisotropic self-assembly, while its integration with lithographic techniques, particularly soft lithography, enables the incorporation of dewetting-growth micro- and nanostructures into advanced optoelectronic devices. 
The presence of an initial nucleation site enables the subsequent growth of a MAPbBr$_3$ crystal directly from the precursor solution on the ITO substrate, ensuring high crystallographic quality together with control over the overall position and dimensions of the resulting crystal ( Fig.\ref{fig:FigDewetting} b).
 Solution-processed growth has previously been reported for arbitrarily shaped micro- and nanostructures with thicknesses in the 500 nm -- 10 $\mu$m range. Here, we extend dewetting-driven bottom-up lithographic approach to the realization of a macroscale perovskite crystal. The resulting MAPbBr$_3$ crystals exhibits thicknesses from 100 to 200 $\mu$m and lateral dimensions exceeding 500 $\mu$m (Figs.\ref{fig:FigDewetting} c,d). The thickness of each crystal has been directly measured using a stereomicroscope (Zeiss, Stemi 305). Notably, each crystal exhibits good large-area structural quality and spatially uniform optoelectronic properties. This structural uniformity enables direct integration with patterned conductive substrates, and thus with the corresponding top and bottom electrical contacts, a key advantage for scalable device fabrication. The quality of the crystal is demonstrated by the spatially-resolved photo-luminescence spectra (Zeiss, LSM980 confocal in  Fig.\ref{fig:FigDewetting} e).

%\section{Device Fabrication}
%\label{sec:device}

A vertical device has been fabricated where the contacts are on opposite faces of the micro-crystal. The  patterned ITO glass (1.1  mm-thick Kintec glass, \SI{1}{\ohm}/square) described in the previous section is the bottom contact of the device. The top ITO contact is deposited by pulsed DC sputtering (KENOSISTEC S.R.L., KS 400 In-Line) 
using a mask with dimensions of 0.2 mm $\times$ 0.5 mm, which is the nominal active area. The deposition parameters are: Pulsed DC Frequency = 30 kHz, Pressure = 1.1 $\times$ 10$^{-3}$ {11} mbar, Ar mass flow = 40 sccm, Power Density = 0.43 W/cm$^{-2}$, Thickness: 180 nm.  

For the results presented here, two devices have been used. 
The first device, with a crystal of 100 $\mu m$ thickness, has been optically characterized in order to extract the main material parameters (Sec. ~\ref{sec:basic_characterization}).
The second device, with a crystal of 150 $\mu m$ thickness, has been exposed to the electron beams (Sec ~\ref{sec:results}).
The crystal quality for both device measured by the
spatially-resolved photo-luminescence spectra is similar.   

\section { Basic electrical characterization}
\label{sec:basic_characterization}
\begin{figure}[ht]
\centering
\includegraphics[ height=1.6in]{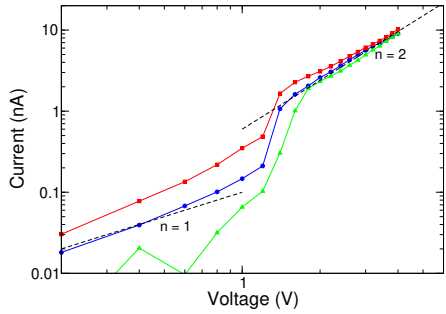}
  \caption{Dark IV characteristics, measured at different times:  the red (blue) points have been taken 12 (18) days after the green points. The dashed lines indicate slopes of 1 and 2. Note that all curves converge to a quadratic voltage dependence above roughly 2 V. }
  \label{fig:IV}
\end{figure}

IV curves in dark conditions have been measured to extract basic material parameters. The measurements have been performed by step wise changing the bias voltage and stabilizing the current over 20 minutes. Some of the resulting IV characteristics, measured at different times, are shown in Fig.~\ref{fig:IV}.
The data has been analyzed based on the space-charge limited current (SCLC) model \cite{LeCorre2021,Liu2019}, motivated by the fact that all measured curves converge to a characteristic with slope of 2 at voltages above 2 V.
From the quadratic part of the IV we extract a carrier mobility of 
%1.12 cm$^2$/Vs was for 200 um
0.47 cm$^2$/Vs, using a relative permittivity of 43.2 as found in \cite{Testa2024}.
From the transition between the regions of high and low slope, we extract a trap-filled limit voltage of approximately 1.4 V, which results in a trap density of 3$\times10^{11}$ cm$^{-3}$.
%1.7$\times10^{11}$ cm$^{-3}$. was for 200 um

\begin{figure}[bt]
\centering
\includegraphics[width=0.95\columnwidth]{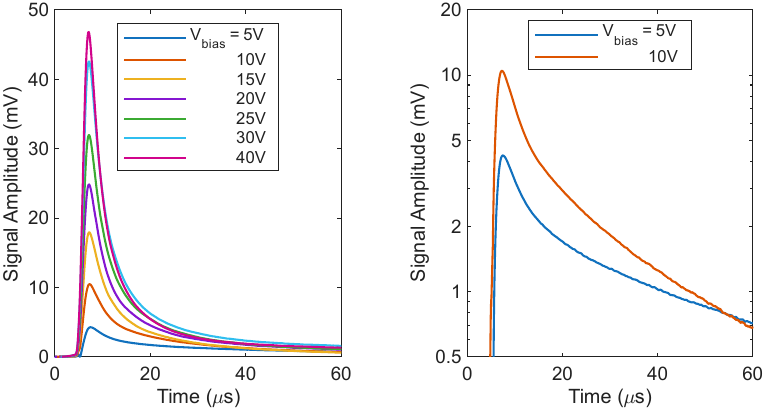}
  \caption{Signal measured with pulsed green laser illumination as a function of applied voltage in linear scale (left) and in logarithmic scale (right).}
  \label{fig:Figure_laser_decay}
\end{figure}

The device was illuminated with a focused pulsed laser diode (Acculase), providing light pulses of 2~$\mu$s width at 520~nm wavelength. At this wavelength, the generation can be considered confined in a thin surface layer of the detector. The photo-generated current was converted into a voltage with a 10-kOhm resistor, and the generated voltage was acquired with a digital storage oscilloscope (Tektronix MDO3102). A positive bias voltage was applied to the electrode hit by laser light in a 2-ms time window with a repetition rate of 10 Hz. The output signals obtained at different bias voltages are shown in Fig.~\ref{fig:Figure_laser_decay} (left). 
The decay could be fitted with a double exponential decay, featuring a short and a long time constant, as can be observed in Fig.~\ref{fig:Figure_laser_decay} (right). The first time constant, of the order of a few microseconds, is connected with the capture time constant of the trapping levels, while the second one, of the order of 10s of microseconds, is related to the release time constant and to the lifetime of holes in the semiconductor \cite{Streetman1966}. 

\begin{figure}[bt]
\centering
\includegraphics[width=0.95\columnwidth]{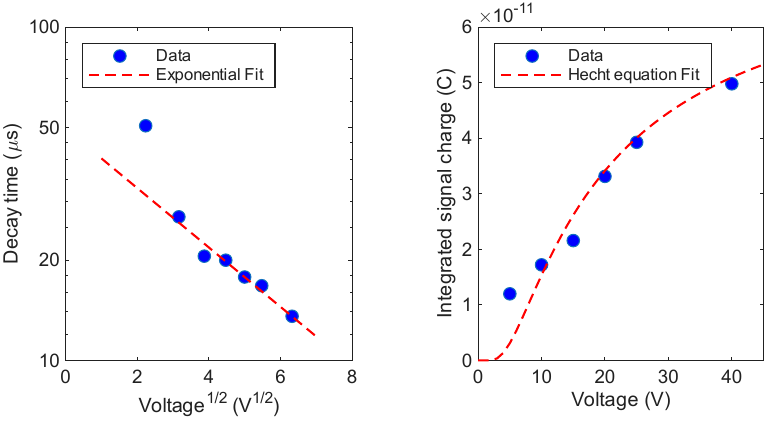}
  \caption{Decay time constant as a function of applied voltage (left). Collected charge as a function of applied voltage (right).}
  \label{fig:Figure_laser_int_tau}
\end{figure}
The decay time represented by the long-time constant is shown in Fig.~\ref{fig:Figure_laser_int_tau} (left). The linear dependence of decay time as a function of the square root of voltage indicates a Poole-Frenkel dependence of emission time constant \cite{Mitrofanov2004}. If compared to the theoretical value, the Poole-Frenkel beta  coefficient is larger, which may indicate the presence of large ionic built-in fields that locally enhance barrier lowering \cite{Alvarez2023}. The collected charge per laser pulse, calculated from the integrated signal, is plotted in Fig.~\ref{fig:Figure_laser_int_tau} (right) as a function of applied voltage. The curve was fitted with a modified Hecht equation, providing a mobility lifetime product of $6.25\times10^{-6}$~cm$^2$/V. 
If an average lifetime ranging from 15 to 50~$\mu$s is considered, a mobility between 0.12 and 0.42~cm$^2$/Vs is obtained, which is in line with the value estimated from the IV curves.
 A collection efficiency of 15.8\% at 5~V, which increases with voltage up to 65\% at 40~V, can be estimated from the data in Fig.~\ref{fig:Figure_laser_int_tau} (right).

% ################################ Section 1 ####################################
\section{Readout Electronics}
\label{sec:electronics}

\begin{figure*}[th]
  \centering
  \includegraphics[width=0.65\linewidth]{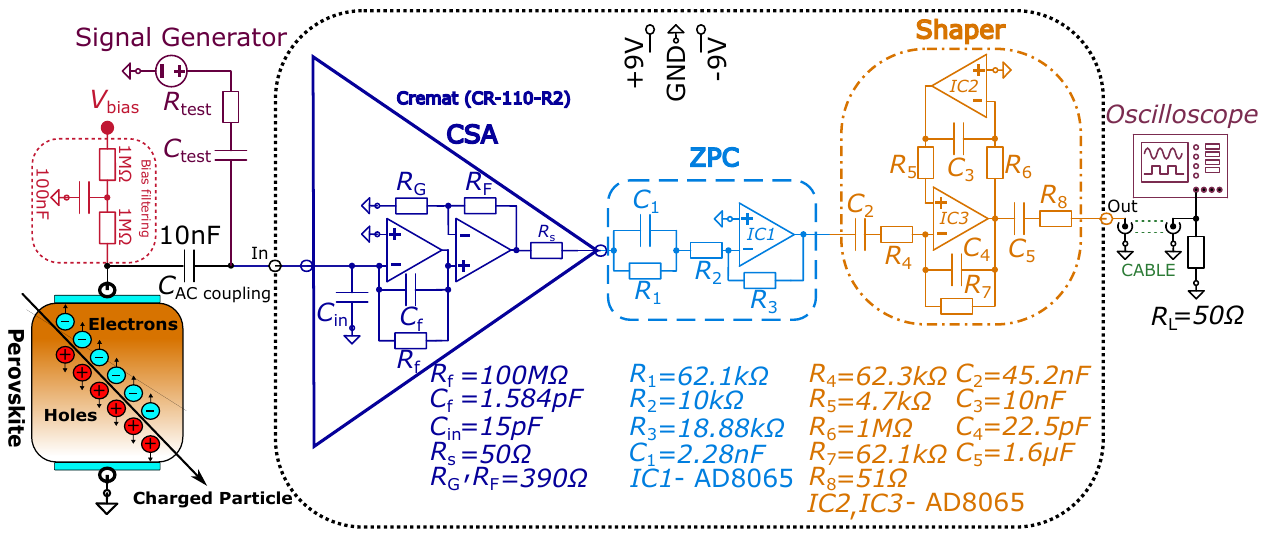}
  \caption{Schematic of the readout electronics chain for the perovskite detector, including biasing and AC coupling, the CSA, pole--zero cancellation, the shaping amplifier, and oscilloscope readout.}
  \label{fig:schematic}
\end{figure*}

Dedicated readout electronics, shown in Fig.~\ref{fig:schematic}, were developed 
and tuned to the characteristics of the signals generated by the perovskite detectors. Unlike conventional semiconductor detectors, perovskite detectors are characterized by relatively slow charge-collection dynamics. To maximize the signal-to-noise ratio and improve the rate capability, the readout chain was therefore optimized for these slower signals.

In a 1.4~mm-thick MAPbBr$_3$ crystal exposed to minimum-ionizing particles (MIPs), the detector current exhibits an exponential decay with a typical time constant of $\mathcal{O}(10)\,\mu$s~\cite{Testa2024}, and similar timing characteristics are observed in the thin devices presented in this work, as discussed in Section III. This behavior requires a dedicated readout chain capable of efficiently processing long detector-current pulses while maintaining low electronic noise. For this reason, for the results presented in this work, a dedicated PCB was designed  and implemented with an adjustable shaping time in the range $\mathcal{O}(1)$--$\mathcal{O}(10)\,\mu$s. This approach is consistent with recent results on perovskite radiation detectors, where readout electronics specifically matched to the detector transport properties, with adjustable shaping/peaking time and reduced parasitic capacitance, were shown to markedly improve the signal-to-noise ratio and overall spectroscopic performance~\cite{Zhang2024}.

The system consists of three main stages: a charge-sensitive amplifier (CSA), a pole--zero cancellation (PZC) network, and a CR--RC shaping amplifier. The shaping amplifier ensures full charge integration while maintaining an acceptable recovery time between events. The PZC stage follows the CSA and compensates for the exponential tail produced by the feedback network, thereby restoring the baseline more rapidly and reducing the probability of pulse pile-up. The shaping stage provides a semi-Gaussian response, which optimizes the signal-to-noise ratio and defines the pulse width. A shaping time constant of \SI{1.4}{\micro\second} was used. The front-end electronics schematic was simulated in LTspice (Ver. 17.2.4)~\cite{LTspice_ADI} and experimentally calibrated using step-voltage injection to determine the nominal gain. In addition, the front-end electronics ballistic deficit, i.e. the pulse-height reduction due to finite charge-collection time relative to the effective integration or measurement time of the readout chain, was measured for exponential input currents with $\tau=\SIrange{3}{105}{\micro\second}$.
The perovskite detector was biased through a filtered bias network  (Fig.~\ref{fig:schematic}), and the signal capacitively coupled to the CSA. The bias circuit provides stable polarization while reducing low-frequency noise contributions from cables and power supplies.

% ################################ Section 2 ####################################
\section{Front-end Electronics Gain Calibration}

The gain of the front-end electronics has been evaluated using LTSpice simulations and calibrated against experimental data.
 The gain has been measured by injecting at the input of the pre-amplifier both a very short current pulse  and a  current with a negative exponential decay.

\subsection{Nominal Gain Calibration}

The nominal front-end electronics gain was calibrated by injecting a short test pulse through the known capacitor shown in Fig.~\ref{fig:schematic}, with $C_{\mathrm{test}}=\SI{1.80}{\pico\farad}$ and $R_{\mathrm{test}}=\SI{0}{\ohm}$, using a voltage step in the range $\Delta V=\SIrange{0.10}{1.00}{\volt}$. The corresponding injected charge therefore spans $Q_{\mathrm{inj}}=\SIrange{0.18}{1.80}{\pico\coulomb}$, according to $Q_{\mathrm{inj}} = C_{\mathrm{test}}\Delta\, V$. This allows the gain to be determined as $G = V_{\mathrm{peak}}/Q_{\mathrm{inj}}$.
From the measured peak amplitudes and the known injected charge, the nominal experimental gain was found to be $936\,\mathrm{mV/pC}$. To reproduce this result in simulation, the LTspice model of the Cremat CSA (CR-110-R2) ~\cite{Cremat_CR110R2} was adjusted by changing the feedback capacitance from $C_f=\SI{1.40}{pF}$ to $C_f=\SI{1.58}{pF}$. With this modification, the LTspice model yields a nominal gain of $936\,\mathrm{mV/pC}$, in good agreement with the experimental value.

\subsection{Ballistic Deficit Measurement}
\label{sec:ballistic_deficit}
To evaluate the response of the readout front-end electronics to an exponentially decaying current pulse, and to quantify the expected gain reduction, the detector current is emulated by a simple RC circuit in the input test line driven by an exponential voltage source (Fig.~\ref{fig:schematic}). In that way, choosing properly the value of $R_\mathrm{test}$ and $C_\mathrm{test}$, a current input with exponential decay is injected at the input of the CSA. This approach provides an analytical expression for the injected current and allows controlled variation of the signal time constant using the signal generator.
The calibration setup consists of an exponential voltage source $V_0(t)$ applied to a resistor, followed by a capacitor $C_\mathrm{test}$ connected to the input of the CSA where  decay time and charge can be controlled only by the signal generator.
When the generator exponential voltage decay time $\tau_g$ is much faster than $\tau_c=R_\mathrm{test}C_\mathrm{test}$ ,
$\tau_g \ll \tau_c$, the circuit generates an approximately pure exponentially decaying current pulse at the CSA input : $i(t) \approx \frac{V_0}{R_\mathrm{test}} e^{-t/\tau_g}$,
where  $i(t)$ is the current injected into the CSA input, $V_0$ is the initial amplitude of the generator voltage.
This method was practical and effective for calibration with exponentially decaying detector-equivalent signals, enabling the study of ballistic deficit and pulse-shaping behavior of the CSA under detector-like current excitation. By varying $\tau_g$ over a suitable range, different charge-collection times can be emulated, and the front-end electronics ballistic-deficit response can be measured as a function of signal time constant. 

\begin{figure}[H]
  \centering
  \includegraphics[ height=1.7in]
{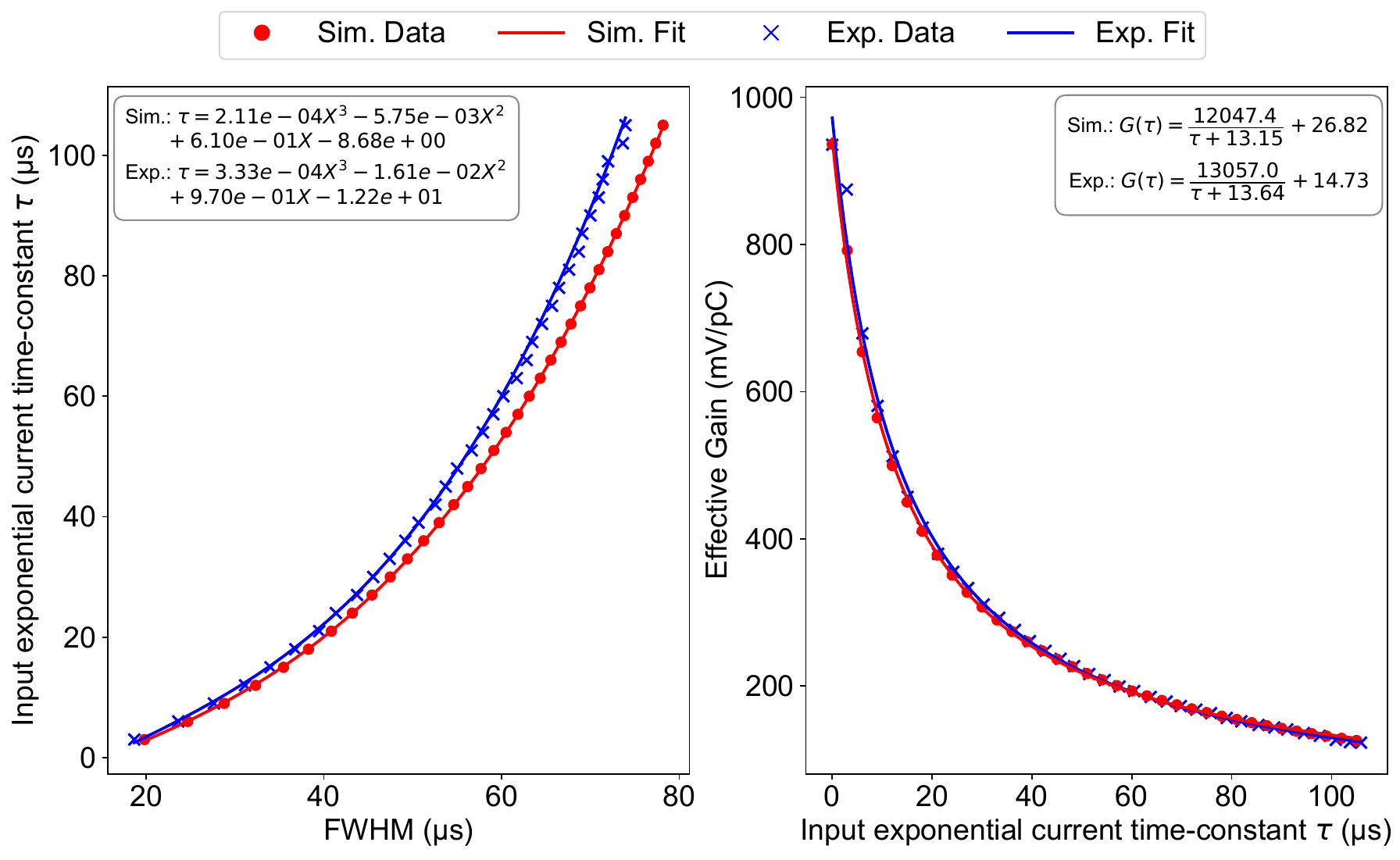}
  \caption{Left: Dependence of the input exponential current time constant, $\tau$, on the shaper-output FWHM, obtained from LTspice simulation and experimental measurements. Right: Effective gain as a function of the input exponential current time constant, $\tau$, obtained from LTspice simulation and experimental measurements.}
\label{fig:fwhm_gain_vs_tau}
\end{figure}

In order calibrate the front-end electronics under ballistic-deficit conditions, an RC calibration network was used (Fig.~\ref{fig:schematic}) with $R_\mathrm{test}=\SI{10}{M\ohm}$ and $C_\mathrm{test}=\SI{1}{nF}$. The time constant was scanned over $\tau=3,6,\ldots,\SI{105}{\micro\second}$.
These pulses were injected while keeping the total injected charge approximately constant ($Q\approx\SI{1.05}{pC}$).
After analyzing the response of the front-end electronics, the effective gain was evaluated as a function of the detector-current decay time constant, $\tau$, which is equivalent to $\tau_g$. Fig.~\ref{fig:fwhm_gain_vs_tau}, right, shows the effective gain obtained from both LTspice simulation and experimental measurements. The effective gain decreases monotonically as $\tau$ increases, indicating that slower detector-current pulses produce a smaller response at the front-end electronics output. This behavior occurs because, for longer detector-current decay times, the same input charge is distributed over a longer time interval, which reduces the maximum amplitude of the shaped output signal.

The data are well described by an empirical shifted-hyperbolic fitting function of the form $G(\tau) = \frac{A}{\tau + B} + C.$ The close agreement between the experimental data and the LTspice simulation confirms that the front-end electronics model accurately reproduces the observed gain dependence on the temporal response of the detector current.
To estimate the detector-current decay time constant, $\tau$, a calibration relationship between the full width at half maximum (FWHM) of the front-end electronics shaper output signal and the input-current time constant is used. 
%\begin{figure}[H]
%  \centering
%  \includegraphics[width=\linewidth]{figures/FWHM_vs_tau_SH.pdf}
%  \caption{Dependence of the shaper-output FWHM on $\tau$ obtained from LTspice simulation and experimental measurements.}
%  \label{fig:fwhm_vs_tau}
%\end{figure}
As shown in %Fig.~\ref{fig:fwhm_vs_tau},
Fig.~\ref{fig:fwhm_gain_vs_tau} left,
both the LTspice simulation and experimental data exhibit a clear monotonic dependence of the shaper-output FWHM on $\tau$. Therefore, the measured FWHM of the shaped output pulse can be used as an estimator of the original detector-current time constant. This approach is stable under noisy conditions, since the FWHM of the shaped signal can be measured more reliably than directly extracting the decay constant from the detector-current waveform. The calibration curve was obtained by fitting the simulated and experimental datasets with a cubic polynomial function. 
%In general, the dependence can be expressed as
%
%\begin{equation}
%    \tau = a \cdot \mathrm{FWHM}^{3} + b \cdot \mathrm{FWHM}^{2} + c %\cdot \mathrm{FWHM} + d,
%    \label{eq:cubic_fwhm_tau}
%\end{equation}
%where $a$, $b$, $c$, and $d$ are fitting coefficients determined from the calibration data. 
Once the FWHM of the front-end electronics output signal is measured, the corresponding value of $\tau$ can be reconstructed using the experimentally obtained polynomial fit.

\section{Test Beam Results}
\label{sec:results}

\begin{figure}[ht]
\centering
\includegraphics[width=0.8\columnwidth]{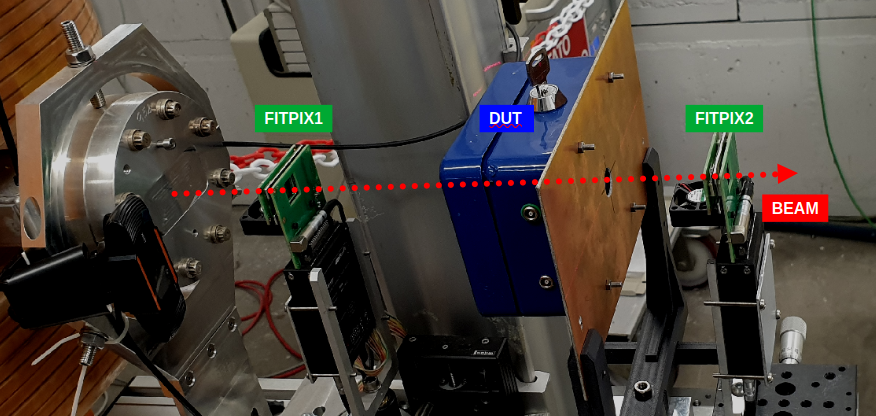}
  \caption{Experimental setup. The thin crystal sample is installed inside the blue box (DUT), that acts as shielding. Beam monitoring device (FITPIX1-2) are also shown. The electron beam (red dotted line) is crossing all the devices before impinging in one EM Calorimeter (not shown) for charge multiplicity measurement. The setup allows to keep under control beam spot size and alignment during the data collection.}
  \label{fig:setup}
\end{figure}

The response of the detector to high-energy electrons has been measured with a dedicated test beam.
The setup of the measurement is shown in Fig.~\ref{fig:setup}.
A 400 MeV electron beam provided by the Beam Test Facility at National Laboratories of Frascati has been used \cite{Foggetta:2021gdg}.  The beam time structure is characterized by bunches of of 10 ns duration with repetition rate of 50 Hz. The multiplicity of electrons inside a bunch can be set from 1 to 10$^4$. Typical transversal width of the beams are of the order of some millimeter, depending on the multiplicity. To characterize our device under test (DUT), whose area is much smaller than the beam spot, it is crucial to evaluate the fraction of the electrons hitting the active area of the crystal.

Two fitpix pixel detectors~\cite{VKraus_2011} (FITPIX1-2 Fig.~\ref{fig:setup}), one after and one in front of the DUT, are used to continuously monitor the position and the spatial distribution of the beam. These two measurements allow to to keep the test crystal aligned with the beam and correctly estimate the fraction of the charge impinging on it.   
A calorimeter placed downstream measures the multiplicity of the electrons in each bunch.

\begin{figure}[ht]
\centering
\includegraphics[height=1.8in]{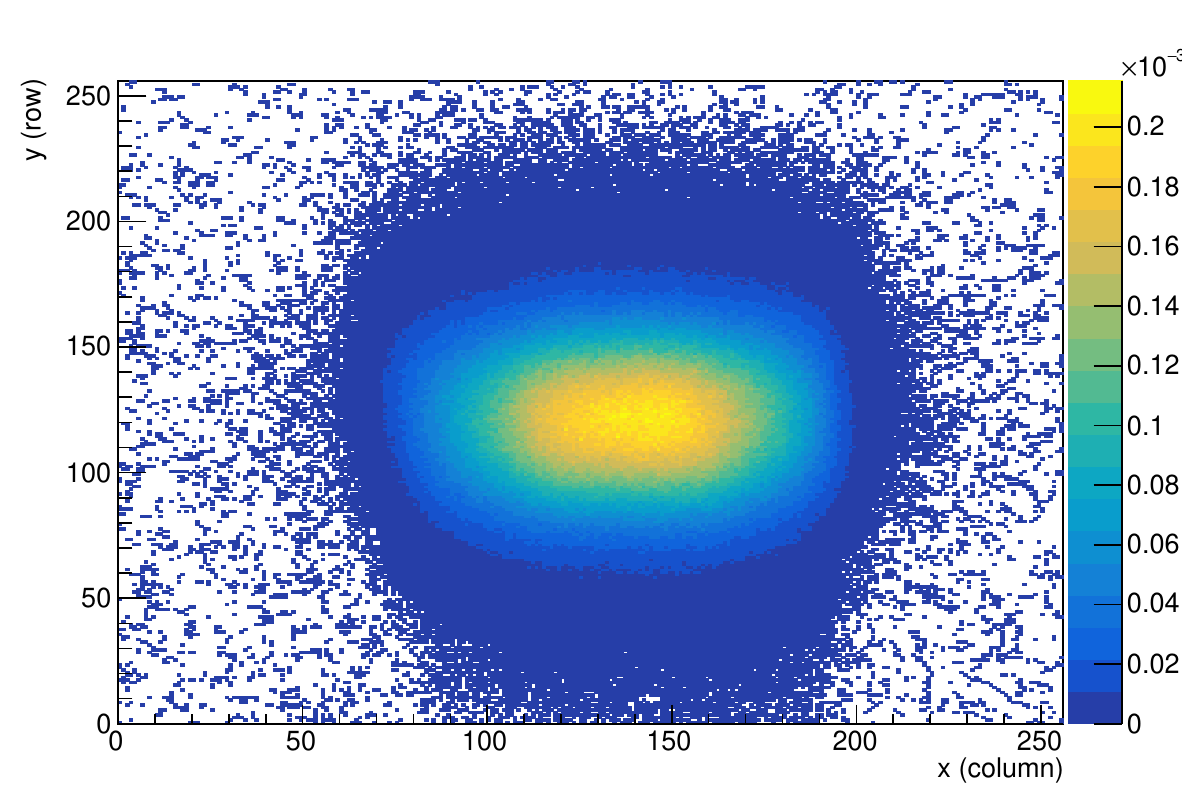}
  \caption{Beam charge distribution measured on the upstream fitpix device for the nominal beam multiplicity of 1000, as  measured by the downstream calorimeter   }
  \label{fig:fitpix1_2D}
\end{figure}

\begin{figure}[ht]
\centering
\includegraphics[width=\columnwidth]{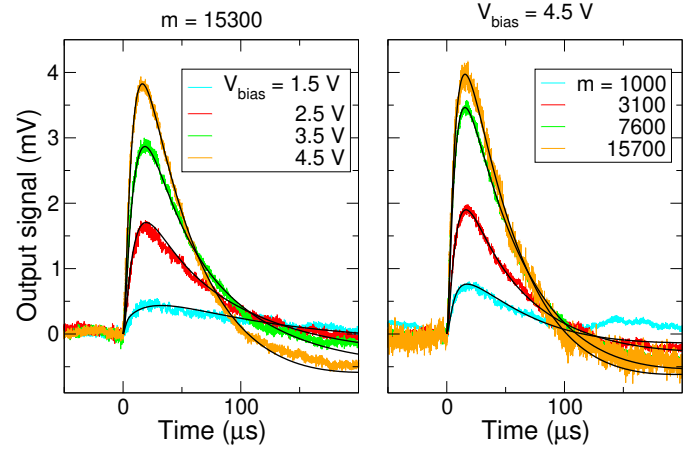}
  \caption{waveforms with at different bias voltages for beam multiplicity of 15300 (left) and at different beam multiplicities with $V_{\mathrm{bias}} = 4.5$ V (right). The nominal beam multiplicities are given, as  measured by the calorimeter. The black lines are the fitted signals. }
  \label{fig:wf_varym}
\end{figure}

The beam spot area on the crystal surface (geometrical acceptance) has been extrapolated run-by-run using the measurements from the fitpix devices, as explained in the following.  The procedure allows to calculate the  effective multiplicity of the beam on the crystal run-by-run.
 Prior to data taking, a calibration procedure to evaluate the beam spot area at the crystal position has been performed by measuring the beam spot distribution using the FITPIX2 in the position of the DUT.
Generally, at the high electron  multiplicities per bunch used in this study, roughly from $10^3$ to $10^4$, the beam distribution is not reproduced by a 2D gaussian distribution (see Fig.~\ref{fig:fitpix1_2D}), as opposed to low multiplicity runs.
The FWHM of the measured distribution on the two fitpix devices in the $x$ and $y$ directions have been considered  to extrapolate geometrically the beam spot at the crystal. The multiple scattering caused by the box containing the device has been taken into account by measuring the spot size at the FITPIX2 position with and without the DUT setup.
Typical values of the extrapolated spot  at the crystal (${\rm FWMH_{crystal}}$) are in the range 4.1-6.5  mm. The probability that an electron hits the crystal in the active area is then calculated as: $ p = f_{\rm {peak}}^{{\rm{fitpix1}}}/({\rm {pitch}}^2)/A \times (\rm {FWMH}_{fitpix1}/FWMH_{crystal})$, where $f_{{\rm{peak}}}^{{\rm{fitpix1}}}$ is the peak of the 2D $f(x,y)$ measured distribution of the beam on the FITPIX1 device, pitch is its pixel pitch equal to 55 $\mu m$,  $A$ is the active area of the micro-crystal as defined in Sec.~\ref{sec:device}.
The calculated probabilities are in the range of 0.3 - 0.6 \% and the resulting effective multiplicity on the active area is 7-36 electrons per bunch.
%It's interesting to note that the fitting function of the  calculated effective multiplicities as a function of the nominal multiplicity is in agreement with the expected electron density at the beam spot center. ({\color{blue} Matthias}).
Fig.~\ref{fig:wf_varym} presents the waveforms recorded at various nominal beam multiplicities.

\begin{figure}[ht]
%\centering
\includegraphics[width=\columnwidth]{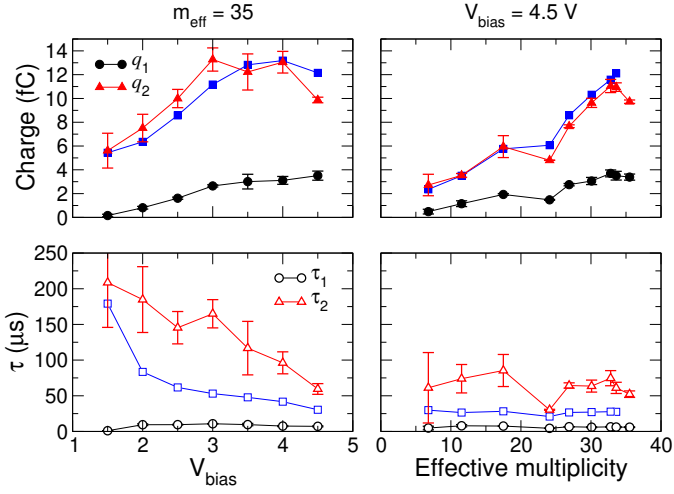}
  \caption{Fitted released charge inside the crystal and decay time constants,  as a function of bias voltage (left) and electron multiplicity per bunch (right). $q_1$, $\tau_1$ and $q_2$, $\tau_2$ correspond to the fast and slow signal component, respectively. The effective multiplicity for the bias dependent data is 35, the applied voltage for the data at different effective multiplicity is is 4.5 V. The blue curves with square symbols show the charge and time constant obtained from the FWHM analysis according to section~\ref{sec:ballistic_deficit}.
  %The dashed black line in the right top figure is a linear fit with a slope of 0.251 fC/e$^-$.
  }
  \label{fig:fit_par}
\end{figure}

Based on the results of the optical characterization presented in section~\ref{sec:basic_characterization}, we have fitted the measured waveforms assuming  for the detector current a double-exponential form given by $I(t) = q_1/\tau_1\times e^{-t/\tau_1}  + q_2/\tau_2\times e^{-t/\tau_2}$, convoluted with the impulse response of the readout electronics (see section~\ref{sec:electronics}).
To reduce the systematic signal component induced by the test-beam facility and the measurement setup, we subtracted a trace acquired at 4.5 V bias with the crystal out of the beam in a time interval from -10 to 30 $\mu$s.
Also, the mean value over 10 $\mu$s before the pulse has been subtracted prior to fitting.
We have included in the fitting procedure the signal from 0 to up to 350 $\mu$s.
From the distribution of the fitting parameters obtained for different fitting intervals we extracted the peak position and the half-width at half maximum. The latter has been used as a measure of parameter uncertainty, and is given in the plots as error bars.
%For the multiplicity-dependent data the fits have a reduced $\chi_\nu^2$ between 1.2 and 1.5, apart from an outlier at $m_{eff}=36$ with $\chi_\nu^2 = 2$.
%The voltage dependent data fits show a $\chi_\nu^2$ between 1 and 1.8.
%Apart from the fit for the lowest bias voltage, all fits show $R^2 > 0.95$, and relative uncertainties less than 6\% in all parameters.
The fitted charge inside the crystal and the decay constants as a function of the effective electron multiplicity per bunch or bias voltage is shown in Fig.~\ref{fig:fit_par}.
The fast component has a time constant of around 6 $\mu$s and contributes roughly 20\% to the total extracted charge.
The time constant of the slow component shows some voltage dependence,  in line with the optical characterization result. The obtained value is slightly larger than that found in \cite{Testa2024} for a bulk single crystal.
For comparison, the results of the analysis based on a single exponential decay model as described in section~\ref{sec:ballistic_deficit} is also shown.
The extracted time constant roughly corresponds to the mean value of the fast and slow time constants, while the charge compares well with that of the slow component obtained from fitting the measured signals.
This confirms that the analysis based on the FWHM of the output signal can be used reliably to reconstruct the extracted charge.
%Note that during the measurements the crystal started to exhibit unstable behavior at bias voltages $\ge$ 4.5 V, so that the last point  in the voltage dependent data might be unreliable.

\begin{figure}[h]
\centering
\includegraphics[width=\columnwidth]{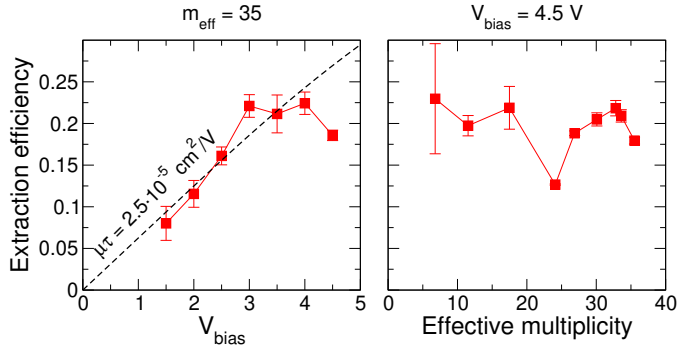}
  \caption{Extraction efficiency as a function of the bias voltage (left) and of effective multiplicity (right). The dashed black line shows the predicted efficiency from the Hecht model, using the extracted mobility of 0.47 cm$^2$/Vs and a mean life time between 50 and 60 $\mu$s.
  %, while the dash-dotted line corresponds to the Hecht model with the same slope as the measured curve.
  }
  \label{fig:effvsmult}
\end{figure}

In order to estimate the charge extraction efficiency, given as the ratio between extracted and generated charge $Q_\mathrm{extr}/Q_\mathrm{gen}$,  we have simulated the energy deposit of 400 MeV electrons inside a 150 $\mu$m thick OMHP crystal  using the Geant4 software~\cite{Geant4}.  The mean released  energy is 75 keV. 
% 0.1 MeV for 200 um
Taking into account the energy to create an electron-hole pair,
calculated using the formula $W = 2E_g + 1.43$ eV~\cite{DEVANATHAN2006637}, where $E_g =
2.2$ eV is the band gap of MAPbBr$_3$, approximately  12864 
$e$-$h$ pairs are generated from the energy released by one electron
passing through the micro-crystal.
% before we had 17520 e-h pairs per electron
The total extracted charge has been calculated as $Q_\mathrm{extr} = q_1 + q_2$.
Fig.~\ref{fig:effvsmult} shows the calculated extraction efficiency both as a function of bias and effective multiplicity.
The voltage dependent efficiency can be compared with the one predicted by the Hecht model.
The measured slope is compatible with the  mobility-lifetime product stated before.
It is worth noticing that  400 MeV electrons behave as  MIPs~\cite{Testa2024} inside a small crystal. Therefore, our device is  expected to be  sensitive also to low energy  electrons which deposit more energy in the crystal. Moreover, since the energy released in a material by a MIP is mostly independent on the type of charged particles,  we expect sensitivity  to protons as well.  

%\newpage
\section{Conclusions and Discussion}
We report for the first time a device based on a 150 $\mu$m thick  OMHP single crystal, able to detect  high energy electrons of $ \sim 400$ MeV.
The device response is  linear within a variation of the incident electron flux of about a factor of 4. Assuming a nominal active area of 0.5$\times$0.2 mm$^2$, the extraction efficiency at 4.5 V bias is around 19\%. At that voltage bias, the detector time constant is around 60 $\mu$s, driven by carrier life-time.
%which is an improvement with respect to the previous measurement on a 1 mm thick crystal~\cite{Testa2024}. 
The extraction efficiency and the rise time measured using high energy electrons are in reasonable agreement with those measured using visible light, with similar voltage bias.
It is worth noting that the efficiency using visible light increase up to 65\% at 40 V. This was possible using a bias voltage with a repetition rate, which mitigates the well-known response instabilities at high voltage bias, likely related to ion migration~\cite{Flannery2025, Bao2026,Tsarev2025}.
This technique, while not suitable for test beam environment, shows the potential to increase the efficiency by increasing the voltage bias.
Further improvements both to increase the extraction efficiency and to improve the timing performance
are expected by increasing the voltage bias using a guard-ring~\cite{guardring} and 
using other perovksites like CsPbBr$_3$ which show better stability at higher voltage bias~\cite{10238716}.
In the future, efforts will be addressed to grow single perovskite crystals with larger area directly on pixelated electronic chips for imaging and monitoring applications~\cite{Campanelli2024}.

\section*{Acknowledgments}
%This should be a simple paragraph before the References to thank those individuals and institutions who have supported your work on this article.
%Funded by the European Union under NextGenerationEU. PRIN 2022 Prot. n. 2022LWHCWY.
We warmly thank the BTF Staff, namely Eleonora Diociauti and Clara Taruggi for their excellent beam setups and scientific support, %. We also extend our gratitude to
the DA$\phi$NE operators and LINAC service for maintaining great uptime in BTF injections.
We deeply thank  Maurizio Gatta and Matteo Beretta of the LNF electronic service for the design and production of the PCB. % and the technical support. 
{\appendix[Derivation of modified Hecht equation]
We calculate the electron and hole current in a metal-semiconductor-metal structure under the following assumptions: (1) the electric field $E$ is constant and diffusion can be neglected; (2) the contact work functions are nearer to the conduction band edge so that $p \ll n$ and trap-assisted Shockley-Read-Hall recombination is given by $R=p/\tau_p$; (3) carrier generation happens at the positive contact in form of a delta-pulse so that holes have to travel through the semiconductor. We then have to solve
\begin{eqnarray}
    \frac{\partial n}{\partial t} - \mu_nE\frac{\partial n}{\partial x} +\frac{p}{\tau_p} & =&  N_0\delta(x)\delta(t) \label{eq:el_cont} \\
    \frac{\partial p}{\partial t} + \mu_pE\frac{\partial p}{\partial x} +\frac{p}{\tau_p} & =&  N_0\delta(x)\delta(t),\label{eq:hole_cont}
\end{eqnarray}
where $N_0$ is the number of $e$-$h$ pairs generated by the pulse.
The electric field is $E=V/L$ where $V$ is the applied bias and $L$ the semiconductor thickness.
$\delta(x)$ is the usual Dirac delta distribution.
After Laplace-transforming eq.~\eqref{eq:hole_cont} from time domain to Laplace domain with coordinate $s$ we obtain
\begin{equation}
    \frac{\mathrm{d}P}{\mathrm{d}x} + \lambda_p(1+\tau_ps)P = 0
\end{equation}
with $\lambda_p=1/(\mu_p\tau_pE)$ and boundary condition $\mu_pEP(0,s)=N_0$.
The solution is
\begin{equation}
    P(x,s) = \frac{N_0}{\mu_pE}e^{-\lambda_p(1+\tau_ps)x}. \label{eq:pdens}
\end{equation}
Further, we write the electron density as $n_0+n(x,t)$, where $n_0$ is the equilibrium density.
After Laplace transformation of eq.~\eqref{eq:el_cont} for the light-induced electron density, and using \eqref{eq:pdens}, we get 
%\begin{equation}
\begin{align}
    N(x,s)  = &\frac{N_0\lambda_p}{\mu_nE}\frac{e^{-\lambda_p(1+\tau_ps)x}}{\lambda_p+\tau_p(\lambda_n+\lambda_p)s} \nonumber \\
     & \times\left\{e^{-\left[\lambda_p+\tau_p(\lambda_n+\lambda_p)s\right](L-x)} - 1\right\}
\end{align}
%\end{equation}
The drift current density becomes
\begin{equation}
    J(x,s)=e\mu_pEP(x,s)+e\mu_nEN(x,s),
\end{equation}
where $e$ is the elementary charge.
The total extracted charge is given by
\begin{equation}
  Q=\lim_{t\rightarrow\infty}\int_0^tj(x,\tau)\mathrm{d}\tau=\lim_{s\rightarrow 0} J(x,s)
\end{equation}
by virtue of the final value theorem and of the Laplace transform of the integral.
The final result becomes
\begin{equation}
    Q=eN_0e^{-\lambda_pL}=eN_0e^{-\frac{L^2}{\mu_p\tau_pV}}
\end{equation}
The same result is obtained when calculating the contact current via the Ramo-Shockley theorem.
Also, the same procedure applied to a pulse with spatially constant generation, $G=N_0/L\delta(t)$, leads to the usual Hecht equation.

}

%Use $\backslash${\tt{appendix}} if you have a single appendix:
%Do not use $\backslash${\tt{section}} anymore after $\backslash${\tt{appendix}}, only $\backslash${\tt{section*}}.
%If you have multiple appendixes use $\backslash${\tt{appendices}} then use $\backslash${\tt{section}} to start each appendix.
%You must declare a $\backslash${\tt{section}} before using any $\backslash${\tt{subsection}} or using $\backslash${\tt{label}} ($\backslash${\tt{appendices}} by itself
% starts a section numbered zero.)}


\bibliography{cas-refs}

@article{LeCorre2021,
author = {Le Corre, Vincent M. and Duijnstee, Elisabeth A. and El Tambouli, Omar and Ball, James M. and Snaith, Henry J. and Lim, Jongchul and Koster, L. Jan Anton},
title = {Revealing Charge Carrier Mobility and Defect Densities in Metal Halide Perovskites via Space-Charge-Limited Current Measurements},
journal = {ACS Energy Letters},
volume = {6},
number = {3},
pages = {1087-1094},
year = {2021},
doi = {10.1021/acsenergylett.0c02599}
}

@article{Liu2019,
title = {Low-temperature-gradient crystallization for multi-inch high-quality perovskite single crystals for record performance photodetectors},
journal = {Materials Today},
volume = {22},
pages = {67-75},
year = {2019},
issn = {1369-7021},
doi = {https://doi.org/10.1016/j.mattod.2018.04.002},
url = {https://www.sciencedirect.com/science/article/pii/S1369702118301172},
author = {Yucheng Liu and Yunxia Zhang and Zhou Yang and Jiangshan Feng and Zhuo Xu and Qingxian Li and Mingxin Hu and Haochen Ye and Xu Zhang and Ming Liu and Kui Zhao and Shengzhong(Frank) Liu}
}

@article{Sakhatskyi2025,
author = {Sakhatskyi, Kostiantyn and Bhardwaj, Aditya and Matt, Gebhard J. and Yakunin, Sergii and Kovalenko, Maksym V.},
title = {A Decade of Lead Halide Perovskites for Direct-Conversion X-ray and Gamma Detection: Technology Readiness Level and Challenges},
journal = {Advanced Materials},
volume = {37},
number = {27},
pages = {2418465},
doi = {https://doi.org/10.1002/adma.202418465},
url = {https://advanced.onlinelibrary.wiley.com/doi/abs/10.1002/adma.202418465},
eprint = {https://advanced.onlinelibrary.wiley.com/doi/pdf/10.1002/adma.202418465},
year = {2025}
}

@article{Yihui2022,
author = {Yihui He and  Ido Hadar and Mercouri G. Kanatzidis },
title = {Detecting ionizing radiation using halide perovskite semiconductors processed through solution and alternative methods},
journal = {Nature Photonics},
volume = {16},
pages = {14-26},
year = {2022},
 doi = {10.1038/s41566-021-00909-5}
}

@article{Liu2024,
    author = {Liu, Runkai and Li, Feng and Zeng, Fang and Zhao, Rubin and Zheng, Rongkun},
    title = {Halide perovskite x-ray detectors: Fundamentals, progress, and outlook},
    journal = {Applied Physics Reviews},
    volume = {11},
    number = {2},
    pages = {021327},
    year = {2024},
    month = {06},
    issn = {1931-9401},
    doi = {10.1063/5.0198695}
    }

@article{Viola2023,
author = {Viola, Ilenia and Matteocci, Fabio and De Marco, Luisa and Lo Presti, Leonardo and Rizzato, Silvia and Sennato, Simona and Zizzari, Alessandra and Arima, Valentina and De Santis, Antonio and Rovelli, Chiara and Morganti, Silvio and Auf der Maur, Matthias and Testa, Marianna},
title = {Microfluidic-Assisted Growth of Perovskite Single Crystals for Photodetectors},
journal = {Advanced Materials Technologies},
volume = {8},
number = {14},
pages = {2300023},
doi = {https://doi.org/10.1002/admt.202300023},
url = {https://advanced.onlinelibrary.wiley.com/doi/abs/10.1002/admt.202300023},
eprint = {https://advanced.onlinelibrary.wiley.com/doi/pdf/10.1002/admt.202300023},
year = {2023}
}

@article{Deng2015,
author ="Deng, Hui and Dong, Dongdong and Qiao, Keke and Bu, Lingling and Li, Bing and Yang, Dun and Wang, Hong-En and Cheng, Yibing and Zhao, Zhixin and Tang, Jiang and Song, Haisheng",
title  ="Growth{,} patterning and alignment of organolead iodide perovskite nanowires for optoelectronic devices",
journal  ="Nanoscale",
year  ="2015",
volume  ="7",
issue  ="9",
pages  ="4163-4170",
publisher  ="The Royal Society of Chemistry",
doi  ="10.1039/C4NR06982J",
url  ="http://dx.doi.org/10.1039/C4NR06982J",
eprint = {},
year = {2015}
}

@article{Mao2017,
author = {Mao, Jian and Sha, Wei E. I. and Zhang, Hong and Ren, Xingang and Zhuang, Jiaqing and Roy, Vellaisamy A. L. and Wong, Kam Sing and Choy, Wallace C. H.},
title = {Novel Direct Nanopatterning Approach to Fabricate Periodically Nanostructured Perovskite for Optoelectronic Applications},
journal = {Advanced Functional Materials},
volume = {27},
number = {10},
pages = {1606525},
doi = {https://doi.org/10.1002/adfm.201606525},
url = {https://advanced.onlinelibrary.wiley.com/doi/abs/10.1002/adfm.201606525},
eprint = {},
year = {2017}
}

@article{Yang2018,
author = {Yang, Xiaoyu and Wu, Jiang and Liu, Tanghao and Zhu, Rui},
title = {Patterned Perovskites for Optoelectronic Applications},
journal = {Small Methods},
volume = {2},
number = {10},
pages = {1800110},
doi = {https://doi.org/10.1002/smtd.201800110},
url = {https://onlinelibrary.wiley.com/doi/abs/10.1002/smtd.201800110},
eprint = {https://onlinelibrary.wiley.com/doi/pdf/10.1002/smtd.201800110},
year = {2018}
}

@article{Li2022,
author = {Li, Shun-Xin and Xia, Hong and Sun, Xiang-Chao and An, Yang and Zhu, He and Sun, Hong-Bo},
title = {Curved Photodetectors Based on Perovskite Microwire Arrays via In Situ Conformal Nanoimprinting},
journal = {Advanced Functional Materials},
volume = {32},
number = {29},
pages = {2202277},
doi = {https://doi.org/10.1002/adfm.202202277},
url = {https://advanced.onlinelibrary.wiley.com/doi/abs/10.1002/adfm.202202277},
eprint = {https://advanced.onlinelibrary.wiley.com/doi/pdf/10.1002/adfm.202202277},
year = {2022}
}

@article{GU20232666,
title = {From planar structures to curved optoelectronic devices: The advances of halide perovskite arrays},
journal = {Matter},
volume = {6},
number = {9},
pages = {2666-2696},
year = {2023},
issn = {2590-2385},
doi = {https://doi.org/10.1016/j.matt.2023.05.007},
url = {https://www.sciencedirect.com/science/article/pii/S2590238523002254},
author = {Zhenkun Gu and Yiqiang Zhang and Yingjie Zhao and Qun Xu and Yanlin Song},
}

@article{Huang2023,
author = {Huang, Huaqing and Guo, Linxin and Zhao, Yunbiao and Peng, Shengyuan and Ma, Wenjun and Wang, Xinwei and Xue, Jianming},
title = {Radiation-Tolerant Proton Detector Based on the MAPbBr3 Single Crystal},
journal = {ACS Applied Electronic Materials},
volume = {5},
number = {1},
pages = {381-387},
year = {2023},
doi = {10.1021/acsaelm.2c01406},
URL = { 
https://doi.org/10.1021/acsaelm.2c01406 
},
eprint = { 
        https://doi.org/10.1021/acsaelm.2c01406   
}
}

@article{Afshari2023,
    author = {Afshari, Hadi and Chacon, Sergio A. and Sourabh, Shashi and Byers, Todd A. and Whiteside, Vincent R. and Crawford, Rose and Rout, Bibhudutta and Eperon, Giles E. and Sellers, Ian R.},
    title = {Radiation tolerance and self-healing in triple halide perovskite solar cells},
    journal = {APL Energy},
    volume = {1},
    number = {2},
    pages = {026105},
    year = {2023},
    month = {09},
    issn = {2770-9000},
    doi = {10.1063/5.0158216},
    url = {https://doi.org/10.1063/5.0158216},
    eprint = {https://pubs.aip.org/aip/ape/article-pdf/doi/10.1063/5.0158216/18118701/026105_1_5.0158216.pdf},
}

@article{Kirmani2024,
author = {Kirmani, A.R. and  Byers, T.A. and  Ni, Z. et al.},
title = {Unraveling radiation damage and healing mechanisms in halide perovskites using energy-tuned dual irradiation dosing.},
journal = {Nature Communication},
Volume = {15},
pages = {696},
year = {2024},
doi ={https://doi.org/10.1038/s41467-024-44876-1}
}

@article{Yin2026,
author = {Yin, H. and  Wu, H. and  Zhang, Y. et al. },
title = {Highly sensitive and stable perovskite detector for ultrahigh-energy radiations via dynamic repair regulation},
journal = {Nature Photonics},
year = {2026}, doi = {https://doi.org/10.1038/s41566-026-01849-8}}

@Article{qubs9040029,
AUTHOR = {Gazis, Nikolaos and Gazis, Evangelos},
TITLE = {Review of VHEE Beam Energy Evolution for FLASH Radiation Therapy Under Ultra-High Dose Rate (UHDR) Dosimetry},
JOURNAL = {Quantum Beam Science},
VOLUME = {9},
YEAR = {2025},
NUMBER = {4},
ARTICLE-NUMBER = {29},
URL = {https://www.mdpi.com/2412-382X/9/4/29},
ISSN = {2412-382X},
DOI = {10.3390/qubs9040029}
}

@article{guardring,
    author = {Pan, Lei and Pandey, Indra Raj and Liu, Zhifu and Peters, John A. and Chung, Duck Young and Hansson, Conny and Wessels, Bruce W. and Miceli, Antonino and Kanatzidis, Mercouri G.},
    title = {Study of perovskite CsPbBr3 detector polarization and its mitigation with ultrahigh x-ray flux},
    journal = {Journal of Applied Physics},
    volume = {133},
    number = {19},
    pages = {194502},
    year = {2023},
    month = {05},
    issn = {0021-8979},
    doi = {10.1063/5.0151902},
    url = {https://doi.org/10.1063/5.0151902},
    eprint = {https://pubs.aip.org/aip/jap/article-pdf/doi/10.1063/5.0151902/17568799/194502_1_5.0151902.pdf},
}

@article{Geant4,
title = {Recent developments in Geant4},
journal = {Nuclear Instruments and Methods in Physics Research Section A: Accelerators, Spectrometers, Detectors and Associated Equipment},
volume = {835},
pages = {186-225},
year = {2016},
issn = {0168-9002},
doi = {https://doi.org/10.1016/j.nima.2016.06.125},
url = {https://www.sciencedirect.com/science/article/pii/S0168900216306957},
author = {J. Allison et al }
}

@Article{Testa2024,
author ="Testa, Marianna and De Santis, Antonio and Tinti, Gemma and Paoloni, Alessandro and Papalino, Giuseppe and Felici, Giulietto and Chubinidze, Zaza and Matteocci, Fabio and Auf der Maur, Matthias and Rizzato, Silvia and Lo Presti, Leonardo and Viola, Ilenia and Morganti, Silvio and Rovelli, Chiara",
title  ="Direct detection of minimum ionizing charged particles in a perovskite single crystal detector with single particle sensitivity",
journal  ="Nanoscale",
year  ="2024",
volume  ="16",
issue  ="27",
pages  ="12918-12922",
publisher  ="The Royal Society of Chemistry",
doi  ="10.1039/D4NR01556H"}

@ARTICLE{Bruzzi2023,
author = {Bruzzi, M. and Calis, N. and Enea, N. and Verroi, E.},
title = {Flexible CsPbCl3 inorganic perovskite thin-film detectors for real-time monitoring in protontherapy},
journal = {Front. Phys.},
volume = {11},
year = {2023}
}

@article{Fratelli2024,
author = {Fratelli, Ilaria and Basiricò, Laura and Ciavatti, Andrea and Margotti, Lorenzo and Cepić, Sara and Chiari, Massimo and Fraboni, Beatrice},
title = {Real-Time Radiation Beam Monitoring by Flexible Perovskite Thin Film Arrays},
journal = {Advanced Science},
volume = {11},
number = {40},
pages = {2401124},
doi = {https://doi.org/10.1002/advs.202401124},
url ={https://advanced.onlinelibrary.wiley.com/doi/abs/10.1002/advs.202401124},
year = {2024}
}

@article{Streetman1966,
   author = {B. G. Streetman},
   doi = {10.1063/1.1703175},
   issn = {0021-8979},
   issue = {8},
   journal = {Journal of Applied Physics},
   month = {7},
   pages = {3137-3144},
   publisher = {AIP Publishing},
   title = {Carrier Recombination and Trapping Effects in Transient Photoconductive Decay Measurements},
   volume = {37},
   url = {/aip/jap/article/37/8/3137/3103/Carrier-Recombination-and-Trapping-Effects-in},
   year = {1966}
}

@article{Mitrofanov2004,
   author = {Oleg Mitrofanov and Michael Manfra and Appl Phys Lett and J Appl Phys},
   doi = {10.1063/1.1719264},
   isbn = {202614:30:05},
   issn = {0021-8979},
   issue = {11},
   journal = {Journal of Applied Physics},
   month = {6},
   pages = {6414-6419},
   publisher = {AIP Publishing},
   title = {Poole-Frenkel electron emission from the traps in AlGaN/GaN transistors},
   volume = {95},
   url = {/aip/jap/article/95/11/6414/470973/Poole-Frenkel-electron-emission-from-the-traps-in},
   year = {2004}
}

@article{Alvarez2023,
   author = {Agustin O. Alvarez and Ferdinand Lédée and Marisé García-Batlle and Pilar López-Varo and Eric Gros-Daillon and Javier Mayén Guillén and Jean Marie Verilhac and Thibault Lemercier and Julien Zaccaro and Lluis F. Marsal and Germà Garcia-Belmonte and Osbel Almora},
   doi = {10.1021/ACSPHYSCHEMAU.3C00002},
   issn = {26942445},
   issue = {4},
   journal = {ACS Physical Chemistry Au},
   month = {7},
   pages = {386-393},
   publisher = {American Chemical Society},
   title = {Ionic Field Screening in MAPbBr3 Crystals Revealed from Remnant Sensitivity in X-ray Detection},
   volume = {3},
   url = {/doi/pdf/10.1021/acsphyschemau.3c00002?ref=article_openPDF},
   year = {2023}
}

@article{VKraus_2011,
doi = {10.1088/1748-0221/6/01/C01079},
url = {https://dx.doi.org/10.1088/1748-0221/6/01/C01079},
year = {2011},
month = {jan},
publisher = {},
volume = {6},
number = {01},
pages = {C01079},
author = {V Kraus and  M Holik and  J Jakubek and  M Kroupa and  P Soukup and  Z Vykydal},
title = {FITPix — fast interface for Timepix pixel detectors},
journal = {Journal of Instrumentation}
}

@article{DEVANATHAN2006637,
title = {Signal variance in gamma-ray detectors—A review},
journal = {Nuclear Instruments and Methods in Physics Research Section A: Accelerators, Spectrometers, Detectors and Associated Equipment},
volume = {565},
number = {2},
pages = {637-649},
year = {2006},
issn = {0168-9002},
doi = {https://doi.org/10.1016/j.nima.2006.05.085},
url = {https://www.sciencedirect.com/science/article/pii/S0168900206009089},
author = {R. Devanathan and L.R. Corrales and F. Gao and W.J. Weber}
}

@article{Bao2026,
author = {Bao, Yining and Ma, Tianshu and Zhang, Yuqi and Shi, Luolei and Qin, Linling and Cao, Guoyang and Wang, Changlei and Li, Xiaofeng and Yang, Zhenhai},
title = {Reverse-Bias Breakdown Mechanisms and Mitigation Strategies in Perovskite Cells and Tandems},
journal = {Advanced Functional Materials},
volume = {36},
number = {29},
pages = {e24250},
doi = {https://doi.org/10.1002/adfm.202524250},
url = {https://advanced.onlinelibrary.wiley.com/doi/abs/10.1002/adfm.202524250},
eprint = {https://advanced.onlinelibrary.wiley.com/doi/pdf/10.1002/adfm.202524250},
year = {2026}
}

@article{Tsarev2025,
  author  = {Tsarev, Sergey and others},
  title   = {Mitigating Ion Migration with Alternating Voltage for Stable Perovskite Image Sensors},
  journal = {ACS Applied Materials \& Interfaces},
  year    = {2025},
  volume  = {17},
  number  = {51},
  pages   = {69635--69644},
  doi     = {10.1021/acsami.5c18552}
}

@article{Flannery2025,
  author  = {Flannery, Laura and others},
  title   = {Unlocking the Dynamics of Ion Migration and Voltage Bias Stress Effects through Crystallite Engineering in Metal Halide Perovskites},
  journal = {ACS Omega},
  year    = {2025},
  volume  = {10},
  number  = {20},
  pages   = {20536--20549},
  month   = {May},
  doi     = {10.1021/acsomega.5c01182}
}

@ARTICLE{10238716,
  author={Jin, Tong and Liu, Yuting and Xiong, Yan and Pang, Jincong and Wu, Haodi and Yuan, Shunsheng and Xu, Ling and Zheng, Zhiping and Tang, Jiang and Niu, Guangda},
  journal={IEEE Electron Device Letters}, 
  title={Enhancing High-Voltage Stability of CsPbBr$_3$ Radiation Detectors Through Surface Treatment and Electrode Replacement}, 
  year={2023},
  volume={44},
  number={10},
  pages={1620-1623},
  doi={10.1109/LED.2023.3311473}}

@article{Campanelli2024,
  author  = {Campanelli, R. B. and Gomes, G. S. and Donatti, M. M. and others},
  title   = {Evaluation and synthesis of perovskite crystals as high-Z sensors for hybrid pixel detectors},
  journal = {Scientific Reports},
  year    = {2024},
  volume  = {14},
  pages   = {27430},
  doi     = {10.1038/s41598-024-74384-7}}

@article{Zhang2024,
  author  = {Zhang, Xin and Bai, Ruichen and Fu, Yuhao and Hao, Yingying and Peng, Xinkai and Wang, Jia and Ge, Bangzhi and Liu, Jianxi and Hu, Yongcai and Ouyang, Xiaoping and Jie, Wanqi and Xu, Yadong},
  title   = {High energy resolution {CsPbBr3} alpha particle detector with a full-customized readout application specific integrated circuit},
  journal = {Nature Communications},
  year    = {2024},
  volume  = {15},
  number  = {1},
  pages   = {6333},
  month   = {Jul},
  doi     = {10.1038/s41467-024-50746-7},
  url     = {https://doi.org/10.1038/s41467-024-50746-7}
}

@misc{LTspice_ADI,
  author       = {{Analog Devices}},
  title        = {{LTspice}},
  howpublished = {\url{https://www.analog.com/en/resources/design-tools-and-calculators/ltspice-simulator.html}},
  note         = {Accessed: May 7, 2026}
}

@misc{Cremat_CR110R2,
  author       = {{Cremat Inc.}},
  title        = {{CR-110-R2 Charge Sensitive Preamplifier: Application Guide}},
  howpublished = {\url{https://www.cremat.com/CR-110-R2.pdf}},
  note         = {Accessed: May 7, 2026}
}

@inproceedings{Foggetta:2021gdg,
    author = "Foggetta, Luca and others",
    title = "{The Extended Operative Range of the LNF LINAC and BTF Facilities}",
    booktitle = "{12th International Particle Accelerator Conference~}",
    doi = "10.18429/JACoW-IPAC2021-THPAB113",
    month = "8",
    year = "2021",
    howpublished = {\url{https://inspirehep.net/literature/2019253}},
}
\bibliographystyle{IEEEtran}
%\section{References Section}
%You can use a bibliography generated by BibTeX as a .bbl file.
% The IEEEtran BibTeX style support page is:
 
%\section{Simple References}
%You can manually copy in the resultant .bbl file and set second argument of $\backslash${\tt{begin}} to the number of references
% (used to reserve space for the reference number labels box).

%\begin{thebibliography}{1}
%\bibliographystyle{IEEEtran}

\newpage

\vfill

\end{document}